\documentclass{elsart}

\usepackage{graphics}
\usepackage{graphicx}

\usepackage{amssymb}

\begin{document}

\begin{frontmatter}



\title{Stretched-exponential decay functions from a self-consistent model of dielectric relaxation}


\author[label1]{A. V. Milovanov\corauthref{cor1}}\ead{Alexander.Milovanov@phys.uit.no}
\author[label2]{J. J. Rasmussen}
\author[label1]{K. Rypdal}
\address[label1]{Department of Physics and Technology, University of Troms\o, N-9037 Troms\o, Norway}
\address[label2]{Optics and Plasma Research Department, Ris\o\,National Laboratory,\\ Technical University of Denmark, Building 128, P.O. Box 49, DK-4000 Roskilde, Denmark}

\corauth[cor1]{{On leave from}: Department of Space Plasma Physics, Space Research Institute, Russian Academy of Sciences, 84/32 Profsoyuznaya street, 117997 Moscow, Russia}


\begin{abstract}
There are many materials whose dielectric properties are described by a stretched exponential, the so-called Kohlrausch-Williams-Watts (KWW) relaxation function. Its physical origin and statistical-mechanical foundation have been a matter of debate in the literature. In this paper we suggest a model of dielectric relaxation, which naturally leads to a stretched exponential decay function. Some essential characteristics of the underlying charge conduction mechanisms are considered. A kinetic description of the relaxation and charge transport processes is proposed in terms of equations with time-fractional derivatives. 
\end{abstract}

\end{frontmatter}

\section{Introduction}
A rich variety of materials with molecular or structural disorder have two important properties in common. One is non-exponential, non-Debye character of dielectric relaxation, often described by a stretched exponential, the so-called Kohlrausch-Williams-Watts (KWW) function \cite{Kohlrausch,Williams}  
\begin{equation}
\phi _{\beta} (t) = \exp [-(t/\tau)^\beta] \label{1.1}
\end{equation}
with the exponent $0 < \beta \leqslant 1$ and $\tau$ a constant. This stretched exponential relaxation has been found empirically in various amorphous materials as for instance in many polymers and glass-like materials near the glass transition temperature (for reviews see Refs. \cite{Kaatz} and \cite{Phillips}, and references therein). Some physical models incorporating general features of the stretched exponential dielectric relaxation are discussed in Refs. \cite{Phillips,Montroll,Dishon} where one also finds reviews of experimental dielectric relaxation data.

The other property, often found in disordered insulators and semiconductors, including those doped with electric charge, is universality of ac (alternating-current) conduction, which is expressible in terms of a power-law dependence of the real part of conductivity on frequency (Ref. \cite{Dyre} for review):
\begin{equation}
\sigma^{\prime} (\omega) \propto \omega ^\eta \label{1.2}
\end{equation}
with the exponent $0 \leqslant \eta < 1$. The universality means that the $\eta$ values do not depend on the details of the underlying conducting lattice nor on the microscopic charge transport mechanism operating in the system (i.e., classical barrier crossing for ions and/or quantum mechanical tunneling for electrons). Both $\beta$ and $\eta$ depend on the chemical composition of the material and the absolute temperature.   

The physical background and the statistical mechanical foundation of the above originally phenomenological expressions have been a matter of debate in the literature. Here we mention two hard results in their favor: (i) The KWW function expands into a weighted superposition of the Debye single exponential relaxation functions \cite{Montroll}
\begin{equation}
\phi _{\beta} (t) = \int _0 ^\infty \varrho _\beta (\mu) \exp (-t/\mu) d\mu
\end{equation}
where the weighting function $\varrho _\beta (\mu)$ is expressible in terms of a stable (L\'evy) distribution. Due to this connection with the statistics of stable laws, stretched exponential relaxation functions can be argued to appear naturally, thus being a characteristic property of systems in which the dynamics occur on many time scales. (ii) The power-law form of ac conduction coefficient derives from a model \cite{Gefen}, in which the conduction occurs as a result of random walks of charged particles on a percolating cluster. In this paradigm the universality of ac conduction is rooted in the universality of the percolation transition \cite{PRB01}. Support for this standpoint can be found in the results of Ref. \cite{Jacobs}. The exponent $\eta$ is expressible in terms of the percolation indices \cite{Gefen} or the topological characteristics of the lattice \cite{PRB01}.  

In this study we analyze the properties of dielectric relaxation in disordered solids on an equal footing with the ac conduction properties. In section 2 we demonstrate that the exponents $\beta$ and $\eta$ are related to each other via $\beta = 1 - \eta$. We then suggest simple numerical estimates of the $\eta$ values in a regime when the conduction concentrates on a percolating cluster. In section 3 we formulate a self-consistent model of dielectric relaxation, in which the KWW decay function is obtained from a power-law dependence of ac conduction coefficient on frequency and the basic electrostatic equations. In section 4 we describe the KWW relaxations kinetically. We are led to the issue of kinetic equations with time fractional derivatives and we propose a systematic derivation of the fractional relaxation and fractional diffusion equations from the property of ac universality. We summarize our results in section 5.       

\section{Stretched-exponential relaxation and ac universality}
\subsection{General}
Let a homogeneous, isotropic dielectric be exposed to the external polarizing electric field ${\bf E} = {\bf E} (t, {\bf {r}})$, which we consider as a function of time $t$ and the space coordinate, ${\bf r}$. By homogeneous and isotropic we refer to spatial scales larger than the typical scales of the molecular/structural disorder. Assuming a linear and spatially local response of the material the polarization field at time $t$ at point ${\bf {r}}$ can be written as    
\begin{equation}
{\bf P} (t, {\bf {r}}) = \int _{-\infty}^{+\infty} \chi (t - t^{\prime}) {\bf E} (t^{\prime}, {\bf {r}}) dt^{\prime}\label{2.1} 
\end{equation}
where $\chi (t - t^{\prime})$ is a response or memory function. Causality requires that $\chi(t - t^{\prime}) = 0$ for $t < t^{\prime}$. A Fourier transformed memory function $\chi (t)$ is defined to be the frequency-dependent complex susceptibility of the material, i.e.,  
\begin{equation}
\chi (\omega) = \int _{-\infty}^{+\infty} \chi (t) e^{i\omega t} dt \label{2.2}
\end{equation}
Given $\chi (\omega)$ one defines the frequency-dependent complex dielectric parameter as 
\begin{equation}
\epsilon (\omega)  = 1 + 4\pi \chi (\omega) \label{2.3}
\end{equation}
In a basic theory of the decay of polarization one is interested in the dielectric response to a field which is steady for $t < 0$ and, then, is suddenly removed at time $t=0$. It is then common to write the complex function $\epsilon (\omega)$ as \cite{Constant} 
\begin{equation}
\frac{\epsilon (\omega) - \epsilon (\infty)}{\epsilon (0) - \epsilon (\infty)} = -\int _0^\infty \frac{d\phi (t)}{dt} e^ {i\omega t} dt \label{2.4}
\end{equation}
where $\epsilon (\infty)$ and $\epsilon (0)$ are, respectively, the high and zero frequency limits of $\epsilon (\omega)$ and $\phi (t)$ is the function that describes the decay of polarization after the polarizing field has been removed. Equation~(\ref{2.4}) is essentially an integral equation for the decay function. The decay function can be obtained directly from the magnitude of the polarization response to an electric field with time history as discussed in Ref. \cite{PRB07}. 

\subsection{Connection between relaxation and memory functions}
It is easy to obtain the connection between the functions $\phi (t)$ and $\chi (t)$. We have, with $\chi (\omega) = \left[\epsilon (\omega) - 1\right] / 4\pi$, 
\begin{equation}
\chi (t) = \frac{1}{2\pi}\frac{\epsilon (0) - 1}{4\pi} \int _{-\infty}^{+\infty} \frac{\epsilon (\omega) - 1}{\epsilon (0) - 1} e^{-i\omega t} d\omega \label{2.5}
\end{equation}
In writing Eq.~(\ref{2.5}) we took into account that $\chi (t)$ and $\chi (\omega)$ form a Fourier pair. Assuming that, in the high frequency limit, the value of the dielectric parameter is, approximately, $\epsilon (\infty) = 1$, and combining Eqs.~(\ref{2.4}) and~(\ref{2.5}) we find
\begin{equation}
\chi (t) = -\frac{\ell}{8\pi ^2} \int _{-\infty}^{+\infty} e^{-i\omega t} d\omega \int _{0}^{+\infty} \frac{d\phi (t^{\prime})}{dt^{\prime}} e^{i\omega t^{\prime}} dt^{\prime} \label{2.6}
\end{equation}
where $\ell = \epsilon (0) - 1$ is the dielectric loss strength. Changing order of integration in Eq.~(\ref{2.6}) we write
\begin{equation}
\chi (t) = -\frac{\ell}{8\pi ^2} \int _{0}^{+\infty} \frac{d\phi (t^{\prime})}{dt^{\prime}} dt^{\prime}\int _{-\infty}^{+\infty} e^{-i\omega (t - t^{\prime})} d\omega \label{2.7}  
\end{equation}
Considering that $\int _{-\infty}^{+\infty} e^{-i\omega (t - t^{\prime})} d\omega = 2\pi \delta (t - t^{\prime})$ from Eq.~(\ref{2.7}) it is found that
\begin{equation}
\chi (t) = - \frac{\ell}{4\pi} \frac{d\phi (t)}{dt} \label{2.8}
\end{equation}
Hence the memory function $\chi (t)$ is just the time derivative of the relaxation function $\phi (t)$ (with a proper coefficient in front). Setting $\phi (t) = \exp [-(t/\tau)^\beta]$ in Eq.~(\ref{2.8}) we get
\begin{equation}
\chi (t) = \beta \frac{\ell}{4\pi \tau ^\beta} t^{\beta - 1} \exp [-(t/\tau)^\beta] \label{2.9}
\end{equation}
Note that $\chi (t)$ diverges as $t^{\beta - 1}$ for $t\rightarrow +0$ when $\beta < 1$. 

\subsection{Leading-term approximation}
We now turn to an explicit calculation of the $\sigma ^{\prime} (\omega)$ function from a KWW decay function. In fact, setting $\phi (t) = \exp [-(t/\tau)^\beta]$ and integrating by parts in Eq.~(\ref{2.4}) we find, for the real and imaginary parts of the complex dielectric parameter, 
\begin{equation}
\epsilon ^{\prime} (\omega) - 1 = \ell [1 - \pi z V (z)] \label{2.10a}
\end{equation}  
\begin{equation}
\epsilon ^{\prime\prime} (\omega) = \ell \pi z Q (z) \label{2.10b}
\end{equation}  
where $z = \omega\tau$ is the dimensionless frequency, $u = t/\tau$ the dimensionless time, and $V (z)$ and $Q (z)$ are the standard integrals:
\begin{equation}
V (z) = \frac{1}{\pi} \int _0^\infty \exp (-u ^\beta) \sin (zu) du \label{2.11a}
\end{equation}    
\begin{equation}
Q (z) = \frac{1}{\pi} \int _0^\infty \exp (-u ^\beta) \cos (zu) du \label{2.11b}
\end{equation} 
which play a prominent role in the theory of stable distributions \cite{Montroll}. In the higher frequency range, physically corresponding to a departure from the ``in phase" conductivity response \cite{Jacobs} and the transition to ac charge transport mechanisms, series expansions of these integrals become \cite{Kaatz,Montroll,Series}
\begin{equation}
zV (z) = \frac{1}{\pi} \sum _{n=0}^\infty (-1)^{n} \frac{1}{z^{n\beta}} \frac{\Gamma (n\beta + 1)}{\Gamma (n + 1)} \cos \frac{n\beta\pi}{2} \label{2.12a}
\end{equation}   
\begin{equation}
zQ (z) = \frac{1}{\pi} \sum _{n=1}^\infty (-1)^{n-1} \frac{1}{z^{n\beta}} \frac{\Gamma (n\beta + 1)}{\Gamma (n + 1)} \sin \frac{n\beta\pi}{2} \label{2.12b}
\end{equation}   
From Eqs.~(\ref{2.12a}) and~(\ref{2.12b}) one can see that the expansion of $\pi zQ (z)$ starts from a term which is proportional to $z ^{-\beta}$ and so does the expansion of $1 - \pi z V(z)$. Writing the susceptibility function as    
\begin{equation}
\chi (\omega) = \frac{\ell}{4\pi} [1 - \pi z V (z) + i \pi z Q (z)] \label{2.13}
\end{equation} 
then going for the leading term in the power expansion of $1 - \pi z V (z) + i \pi z Q (z)$ we find
\begin{equation}
\chi (\omega) \propto \omega ^{-\beta} \label{2.14}
\end{equation}   
Hence
\begin{equation}
\epsilon ^{\prime} (\omega) - 1 \propto \omega ^{-\beta} \label{2.14a}
\end{equation} 
\begin{equation}
\epsilon ^{\prime\prime} (\omega) \propto \omega ^{-\beta} \label{2.14b}
\end{equation} 
Noticing that the real part of the frequency-dependent complex conductivity $\sigma (\omega)$ is expressible as
\begin{equation}
\sigma ^{\prime} (\omega) = \omega \epsilon ^{\prime\prime} (\omega) / 4\pi \label{2.15}
\end{equation} 
from Eqs.~(\ref{2.10b}) and~(\ref{2.15}) we get
\begin{equation}
\sigma^{\prime} (\omega) = \frac{\ell}{4\tau} z^2 Q (z) \label{2.16}
\end{equation}   
with the leading term  
\begin{equation}
\sigma^{\prime} (\omega) \propto \omega ^{1 - \beta} \label{2.17}
\end{equation}   
Equation~(\ref{2.17}) reproduces the phenomenological expression in Eq.~(\ref{1.2}) with the power exponent 
\begin{equation}
\eta = 1 - \beta \label{2.17q}
\end{equation}  
If the conduction occurs on a percolating cluster, the $\eta$ values can be obtained from \cite{PRB01}
\begin{equation}
\eta = \frac{\theta + d - d_f}{2+\theta}\label{Eta}
\end{equation} 
Here, $\theta$ is the exponent of anomalous diffusion \cite{Gefen}, which also appears in describing the dc conductivity near percolation, $d$ is the topological (integer) dimension of the ambient real space, and $d_f \leqslant d$ is the Hausdorff fractal dimension of the subset, on which the conduction is concentrated. Using known estimates \cite{Naka} for $\theta$ and $d_f$ one has $\eta \simeq 0$,~0.3, and~0.6 for $d=1$,~2, and~3, respectively. The mean-field result, holding for $d\geq 6$, is $\eta = 1$. From Eqs.~(\ref{2.17q}) and~(\ref{Eta}) it is found that    
\begin{equation}
\beta = \frac{2 - d + d_f}{2+\theta}\label{Beta}
\end{equation} 
yielding $\beta \simeq 1$,~0.7, and~0.4 for $d=1$,~2, and~3. Interestingly, the estimate $\beta \simeq 0.4$, holding for $d=3$, fits the original result due to Kohlrausch (Refs. \cite{Kohlrausch} and \cite{Phillips} for review). This value corresponds to the value $\eta = 1 - \beta \simeq 0.6$, which nicely falls into the range of theoretical arguments \cite{PRB01} and the existing observational evidence \cite{Dyre,Jacobs,Rolla}.

\section{Self-consistent dynamic-relaxation model}
\subsection{Formulation of the model}
Our goal now is to obtain the KWW decay function analytically. We start with a polarization field of the form
\begin{equation}
{\bf P} (t, {\bf {r}}) = {\bf P} (0, {\bf {r}}) + \int _{0}^{+\infty} \chi (t - t^{\prime}) {\bf E} (t^{\prime}, {\bf {r}}) dt^{\prime}\label{3.1} 
\end{equation}
where ${\bf P} (0, {\bf {r}})$ is the initial polarization, and ${\bf E} (t^{\prime}, {\bf {r}})$ is the electric field. We shall assume that there are no external electric fields acting in the system after time $t = 0$, so that ${\bf E} (t^{\prime}, {\bf {r}})$ is essentially the inherent, self-consistent electric field in the bulk of the dielectric due to the electric charges present. 

Let $\rho (t, {\bf {r}})$ be the density of the electric charges at time $t$ at point ${\bf r}$. The function $\rho (t, {\bf {r}})$ is defined as the mean density of the charges in a small volume around ${\bf r}$ such that the highly fluctuating molecular densities are averaged out. If there are no external charges, the density $\rho (t, {\bf {r}})$ is essentially the density of the polarization charges, which we shall denote by $\tilde{\rho} (t, {\bf {r}})$. If the external charges are present in the bulk of the dielectric with the density $\rho _{\rm ext} (t, {\bf {r}})$ then
\begin{equation}
\rho (t, {\bf {r}}) = \tilde{\rho} (t, {\bf {r}}) + \rho _{\rm ext} (t, {\bf {r}}) \label{sum} 
\end{equation}
By external charges we mean charges of external origin, that are alien to the material. These charges may be present due to a charge injection. By allowing for external charges we include a class of doped insulators and semiconductors, which offer a challenging set of fundamental problems, as well as prominent technological applications (Refs. \cite{Jerome,Anta} and references therein). Equation~(\ref{sum}) extends the model in Ref. \cite{PRB07} in which $\rho _{\rm ext} (t, {\bf {r}}) = 0$. 

In the basic theory of dielectrics one writes
\begin{equation}
\nabla\cdot {\bf E} (t, {\bf {r}}) = 4\pi\rho (t, {\bf {r}})
\label{3.2} 
\end{equation}
and
\begin{equation}
\nabla\cdot {\bf P} (t, {\bf {r}}) = -\tilde{\rho} (t, {\bf {r}})
\label{3.3} 
\end{equation}
so that
\begin{equation}
\nabla\cdot {\bf D} (t, {\bf {r}}) = 4\pi\rho _{\rm ext} (t, {\bf {r}})
\label{3.4} 
\end{equation}
where ${\bf D} = {\bf E} + 4\pi {\bf P}$ is the electric displacement in the medium. The density of the polarization currents is defined by 
\begin{equation}
\tilde{{\bf j}} (t, {\bf {r}}) = \frac{\partial}{\partial t} {\bf P} (t, {\bf {r}}) \label{Pcurrent} 
\end{equation}
Performing $\partial / \partial t$ on Eq.~(\ref{3.1}) it is found that
\begin{equation}
\tilde{{\bf j}} (t, {\bf {r}}) = \int _{0}^{+\infty} \tilde{\sigma} (t - t^{\prime}) {\bf E} (t^{\prime}, {\bf {r}}) dt^{\prime} \label{PPcurrent} 
\end{equation} 
where we introduced
\begin{equation}
\tilde{\sigma} (t - t^{\prime}) = \frac{\partial}{\partial t} \chi (t - t^{\prime}) \label{Pconduct} 
\end{equation} 
Note that, due to causality, $\tilde{\sigma} (t - t^{\prime}) = 0$ for $t < t^{\prime}$. 

If external charges are present in the system, they may cause, in addition, their own current, ${\bf j} _{\rm ext} (t, {\bf {r}})$. Before we write an expression for ${\bf j} _{\rm ext} (t, {\bf {r}})$ we address the issue of the microscopic charge transport mechanism: 

It is generally believed that, in relatively poor conductors such as major glasses and polymers, the polarization current is caused by orientational motion of polar molecules or dipoles containing parts of these \cite{Kaatz,Constant}. The external current, in its turn, could be thought of as arising from a migration of charged particles along the underlying stationary molecular distribution. This migration could be mediated by a specific chemical composition of the material, as for instance by the bonding structure in conducting polymers \cite{Jerome}. In porous, nano-crystalline materials, the transport of charge may also show a strong, nonlinear dependence on the charge density and injection, a phenomenon usually explained in terms of trap-filling (Ref. \cite{Anta} and references therein). As a model approximation, here we shall rely on the hypothesis of trap-controlled conduction and diffusion, in which the transport occurs as a result of hopping \cite{Dyre} of charged particles between the localized states. If the hopping has a characteristic time, then the transport is described by a Markovian chain process with a characteristic hopping frequency. In a more general situation there is a distribution of waiting or residence times between the consecutive steps of the motion and the Markovian property is invalidated. The current density is then a flow with memory:   
\begin{equation}
{\bf j} _{\rm ext} (t, {\bf {r}}) = \int _{0}^{+\infty} \sigma _{\rm ext} (t - t^{\prime}) {\bf E} (t^{\prime}, {\bf {r}}) dt^{\prime}\label{3.6} 
\end{equation}  
where $\sigma _{\rm ext} (t - t^{\prime})$ is a memory function which describes the multi-scale trapping and detrapping of the external charges in wide-gap potential wells of the conduction-band level. Due to causality, $\sigma _{\rm ext} (t - t^{\prime}) = 0$ for $t < t^{\prime}$. 

The total current density in the bulk of the material can now be written as
\begin{equation}
{\bf j} (t, {\bf {r}}) = \tilde{{\bf j}} (t, {\bf {r}}) + {\bf j} _{\rm ext} (t, {\bf {r}})\label{3.7-0} 
\end{equation}  
Utilizing Eqs.~(\ref{PPcurrent}) and~(\ref{3.6}) we have  
\begin{equation}
{\bf j} (t, {\bf {r}}) = \int _{0}^{+\infty} \sigma (t - t^{\prime}) {\bf E} (t^{\prime}, {\bf {r}}) dt^{\prime}\label{3.7} 
\end{equation}  
with
\begin{equation}
\sigma (t - t^{\prime}) = \tilde{\sigma} (t - t^{\prime}) + \sigma _{\rm ext} (t - t^{\prime}) \label{FullMemory} 
\end{equation}  
A Fourier transformed $\sigma (t)$ is defined to be the frequency-dependent complex conductivity of the material, i.e., 
\begin{equation}
\sigma (\omega) = \tilde{\sigma} (\omega) + \sigma _{\rm ext} (\omega) \label{FullConduct} 
\end{equation}
where we introduced the partial ac conductivities $\tilde{\sigma} (\omega)$ and $\sigma _{\rm ext} (\omega)$ due to respectively the polarization and external charges. The conservation of the electric charge is expressed by the continuity equation 
\begin{equation}
\frac{\partial}{\partial t} \rho (t, {\bf{r}}) + \nabla\cdot {\bf j} (t, {\bf {r}}) = 0\label{3.11}
\end{equation}
Substituting ${\bf j} (t, {\bf {r}})$ from Eq.~(\ref{3.7}) and taking $\nabla\cdot$ under the time integration we have
\begin{equation}
\frac{\partial}{\partial t} \rho (t, {\bf{r}}) + \int _{0}^{+\infty} \sigma (t - t^{\prime}) \nabla\cdot {\bf E} (t^{\prime}, {\bf {r}}) dt^{\prime} = 0\label{3.12}
\end{equation}
Utilizing Eq.~(\ref{3.2}) we finally arrive at a closed integro-differential equation for the charge density, i.e.,
\begin{equation}
\frac{\partial}{\partial t} \rho (t, {\bf{r}}) + 4\pi \int _{0}^{+\infty} \sigma (t - t^{\prime}) \rho (t^{\prime}, {\bf {r}}) dt^{\prime} = 0\label{3.13}
\end{equation}

\subsection{Stretched exponential relaxation functions}
By Laplace transforming Eq.~(\ref{3.13}) we find 
\begin{equation}
\mathrm{s}\rho (\mathrm{s}, {\bf{r}}) - \rho (0, {\bf{r}})  + 4\pi\sigma (\mathrm{s})\rho (\mathrm{s}, {\bf{r}}) = 0\label{3.13L} 
\end{equation}
where $\rho (0, {\bf{r}})$ is the density of the charges at time $t=0$, and $\sigma (\mathrm{s})$ is the Laplace transform of $\sigma (t)$. We now speculate on the form of the $\sigma (\mathrm{s})$ function: 

There is an increasing belief \cite{Report} that scale-invariance and fractality are hallmarks of chaos and disorder. That such an argument leads to a power-law behavior of ac conduction coefficient was pointed out by Milovanov and Rasmussen \cite{PRB01} who based this on the early work of Gefen et al. \cite{Gefen}. One would expect that, for those time scales on which the dynamics are dictated by the disorder, the partial ac conductivities can be modeled by power-laws, i.e., $\tilde{\sigma} (\omega) \propto \omega ^{\eta _1}$ and $\sigma _{\rm ext} (\omega) \propto \omega ^{\eta _2}$ with some fractional $\eta _1$ and $\eta _2$. Here we assume that $\eta _1 \simeq \eta _2$, i.e., that the two exponents are approximately the same. This assumption refers to universality of ac conduction \cite{Dyre,Jacobs} and indicates that the scaling properties of conduction of polarization charges and external charges are determined by the fractal geometric properties of the material and not by the details of the conduction mechanism. Writing $\tilde{\sigma} (\omega) \propto \omega ^{\eta}$ and $\sigma _{\rm ext} (\omega) \propto \omega ^{\eta}$ with the same power $\eta$ we have, for the total conductivity 
\begin{equation}
\sigma (\mathrm{s}) = \alpha \mathrm{s} ^\eta\label{Total} 
\end{equation}
with $\alpha$ a constant coefficient. This power-law form is just the Laplace version of Eq.~(\ref{1.2}). Note that $\alpha$ may generally depend on the average concentration of the external charges in the conducting domain. 

Separating variables in Eq.~(\ref{3.13L}) we write $\rho (\mathrm{s}, {\bf{r}}) = \phi (\mathrm{s})\psi({\bf{r}})$ with $\psi({\bf{r}}) = \rho (0, {\bf{r}})$ the initial charge-density. Combining Eqs.~(\ref{3.13L}) and~(\ref{Total}) it is found that
\begin{equation}
\phi (\mathrm{s}) = \frac{1}{\mathrm{s} + 4\pi \alpha \mathrm{s} ^\eta}\label{3.14}
\end{equation}
In the time domain, 
\begin{equation}
\phi (t) = \frac{1}{2\pi i} \int _{-i\infty}^{+i\infty} \frac{e^{\mathrm{s} t}}{\mathrm{s} + \tau ^{-\beta}\,\mathrm{s} ^{1-\beta}} d\mathrm{s}\label{3.15T}
\end{equation}
where we introduced the notations $\beta = 1-\eta$ and $\tau ^{-\beta} = 4\pi \alpha$. Equation~(\ref{3.15T}) coincides with the definition of the Mittag-Leffler function $\mathrm{E} _\beta \left[-(t/\tau) ^{\beta}\right]$ (Eq. (B.1) in Appendix B of Ref. \cite{Klafter}). The Mittag-Leffler function has the series expansion
\begin{equation}
\mathrm{E} _\beta \left[-(t/\tau) ^{\beta}\right] = \sum _{n=0}^{\infty} (-1)^n \frac{(t/\tau) ^{n\beta}}{\Gamma (n\beta + 1)}\label{Expan}
\end{equation}
For short times, this expansion goes as a stretched exponential,
\begin{equation}
\mathrm{E} _\beta \left[-(t/\tau) ^{\beta}\right] \approx \exp \left[-(t/\tau)^\beta / \Gamma (\beta + 1)\right]\label{SExpan}
\end{equation} 
This closed analytical form replicates the KWW decay function in Eq.~(\ref{1.1}). Finally, for the charge relaxation by ac charge transport mechanisms,
\begin{equation}
\rho (t, {\bf{r}}) \propto \exp \left[-(t/\tau)^\beta / \Gamma (\beta + 1)\right] \label{Rho}
\end{equation} 
where we omitted the space dependence for simplicity. Limiting cases of expression~(\ref{Rho}) are the following: 

$\tilde \rho (t, {\bf{r}}) \gg \rho _{\rm ext} (t, {\bf{r}})~-$~Equation~(\ref{Rho}) leads to a stretched exponential relaxation of the polarization charges. For many years, theoretical justification of this stretched exponential relaxation regime has been an issue in the theory of dielectric relaxation \cite{Finance}. 

$\rho _{\rm ext} (t, {\bf{r}}) \gg \tilde \rho (t, {\bf{r}})~-$~Equation~(\ref{Rho}) leads to a stretched exponential relaxation of the external charges. This regime may be appropriate for poorly polarizable disordered media when the above condition is satisfied due to a small $\tilde \rho (t, {\bf{r}})$. An observational verification of this regime might constitute an experimental challenge.

\section{Fractional kinetic equations}

The purpose of this section is to describe the KWW relaxations kinetically. We intend to demonstrate that the power-law dependence of the ac conduction coefficient on frequency leads to a fractional extension of the relaxation and diffusion equations, which accommodate fractional-order time derivatives. These offer a suitable analytic formalism to incorporate the features of ac universality and the underlying structural disorder. The discussion below draws on the paradigm of fractional kinetics \cite{Nature}, which finds expanding applications in various fields of research \cite{Hilfer,Sokolov,Rest,UFN}.

\subsection{Fractional relaxation equation}
First, we notice that the power-law $\mathrm{s} ^\eta$ with the fractional $0 < \eta < 1$ is the Laplace transform of the Riemann-Liouville derivative, which is defined through \cite{Oldham} 
\begin{equation}
_{0} \mathrm{D}_t^{\eta} \psi (t, {\bf{r}}) = \frac{1}{\Gamma (1-\eta)} \frac{\partial}{\partial t} \int _{0}^{t} dt^{\prime} \frac{\psi (t^{\prime}, {\bf{r}})}{(t-t^{\prime})^{\eta}}\label{Riemann}
\end{equation}
with $\psi (t, {\bf{r}})$ a function from the class of differintegrable functions. The Riemann-Liouville derivative is a well-defined fractional extension of the ordinary partial time derivative. Setting $\sigma (\mathrm{s}) = \alpha \mathrm{s} ^\eta$ in the dispersion relation~(\ref{3.13L}) and replacing $\mathrm{s} ^\eta$ by $_{0} \mathrm{D}_t^{\eta}$ we write, with $\eta = 1 - \beta$ and $\tau ^{-\beta} = 4\pi \alpha$, 
\begin{equation}
\frac{\partial}{\partial t} \rho (t, {\bf{r}}) = -\tau ^{-\beta}\, {_0} \mathrm{D}_t^{1 - \beta} \rho (t, {\bf{r}})\label{Frelax}
\end{equation}
Equation~(\ref{Frelax}) is the canonical form of the fractional relaxation equation \cite{Nonn}. Applications of this are reviewed in Refs. \cite{Klafter,Sokolov,Rest}. Here we add to the existing knowledge by proposing that relaxations in disordered solids are described by the fractional relaxation equation, provided that the dynamics are self-consistent, and the property of ac universality is verified. 

\subsection{Fractional diffusion equation describing sub-diffusion}
The above analysis applies to length scales much longer than the mean-free paths of charges participating in the ac conduction processes. At length scales comparable to or shorter than these, the dynamics of relaxation should be described kinetically. The key issue is the form of the flow function, which we define as  
\begin{equation}
\textbf{j} (t, {\bf{r}}) = - \int _{0}^{t} \mathcal{D} (t - t^{\prime}) \nabla\rho (t^{\prime}, {\bf{r}}) dt^{\prime}\label{Fick}
\end{equation}
with a memory kernel $\mathcal{D} (t - t^{\prime})$ such that $\mathcal{D} (t - t^{\prime}) = 0$ for $t < t^{\prime}$. When $\mathcal{D} (t - t^{\prime})$ is a delta function, Eq.~(\ref{Fick}) reduces to the well known, Fick's law. A Fourier transformed $\mathcal{D} (t)$ is defined as the frequency-dependent complex diffusion coefficient, $\mathcal{D} (\omega)$. The value of $\mathcal{D} (\omega)$ can be expressed in terms of the ac conduction coefficient as 
\begin{equation}
\mathcal{D} (\omega) = \frac{T}{n e^2 }\sigma (\omega)\label{Einstein} 
\end{equation} 
Here, $e$ denotes the carrier charge, $n$ their number density, and $T$ the absolute temperature. Equation~(\ref{Einstein}) indicates that the properties of charge conduction and diffusion are determined, for each time and frequency scale, by essentially the same collisional properties. For the diffusion on fractals, Eq.~(\ref{Einstein}) can be obtained as a Fourier transform of the average size- and time-scale dependent diffusion coefficient \cite{Gefen}. In the zero-frequency limit, Eq.~(\ref{Einstein}) reduces to the conventional Einstein relation between the diffusion constant and the dc conductivity. 

Combining Eqs.~(\ref{1.2}) and~(\ref{Einstein}) we can propose that, in the frequency range in which the ac conduction coefficient can be modeled by a power law, 
\begin{equation}
\mathcal{D} (\omega) \propto \omega ^{\eta}\label{Power} 
\end{equation}
Such power-law behavior of the frequency-dependent diffusion coefficient has been found in, for instance, stochastic Hamiltonian systems (Ref. \cite{PRE01} and references therein).

By Laplace transforming Eq.~(\ref{Fick}) we get
\begin{equation}
\textbf{j} (\mathrm{s}, {\bf{r}}) = - \mathcal{D} (\mathrm{s}) \nabla\rho (\mathrm{s}, {\bf{r}})\label{3.17} 
\end{equation}
where $\mathcal{D} (\mathrm{s})$ is the Laplace transform of $\mathcal{D} (t)$. When substituted into the continuity Eq.~(\ref{3.11}) this yields  
\begin{equation}
\mathrm{s}\rho (\mathrm{s}, {\bf{r}}) - \rho (0, {\bf{r}})  = \mathcal{D} (\mathrm{s}) \nabla ^2 \rho (\mathrm{s}, {\bf{r}})\label{3.18} 
\end{equation}
Adhering to the power-law form $\sigma (\mathrm{s}) = \alpha \mathrm{s} ^\eta$ from Eq.~(\ref{Einstein}) we have $\mathcal{D} (\mathrm{s}) = \Lambda \mathrm{s} ^{\eta}$ with $\Lambda = \alpha T/ne^2$. Utilizing the scaling $\mathcal{D} (\mathrm{s}) \propto \mathrm{s} ^{\eta}$ in Eq.~(\ref{3.18}) we write
\begin{equation}
\mathrm{s}\rho (\mathrm{s}, {\bf{r}}) - \rho (0, {\bf{r}})  = \mathrm{s} ^{\eta} \nabla ^2 \rho (\mathrm{s}, {\bf{r}})\label{3.18a} 
\end{equation}
where $\Lambda = 1$ for simplicity. In the time domain, Eq.~(\ref{3.18a}) becomes
\begin{equation}
\frac{\partial}{\partial t} \rho (t, {\bf{r}}) = {_0} \mathrm{D}_t^{1-\beta} \nabla^2 \rho (t, {\bf{r}})\label{FD} 
\end{equation}
where we used $\eta = 1-\beta$. Equation~(\ref{FD}) is the canonical form of the fractional diffusion equation describing sub-diffusion \cite{Klafter}, with $\beta$ the fractal dimension in time \cite{PhysicaD}. In various settings, this equation has been derived and discussed in the literature \cite{Report,Klafter,Sokolov,Rest,UFN,PRE01,PhysicaD,Balak,Zaslavsky,Coffey}. 

The characteristic function which is the two sided Fourier transform of $\rho (t, {\bf{r}})$ over the space variable ${\bf{r}}$ satisfies 
\begin{equation}
\frac{\partial}{\partial t} \rho (t, {\bf{k}}) = -{\bf k} ^2 \, {_0} \mathrm{D}_t^{1-\beta} \rho (t, {\bf{k}})\label{Char} 
\end{equation}
Equation~(\ref{Char}) is essentially the fractional relaxation equation in wave-vector space. For the small $\Lambda {\bf k} ^2 t^\beta \lesssim 1$, the characteristic function reduces to a stretched exponential decay function:  
\begin{equation}
\rho (t, {\bf{k}}) \approx \exp \left[-{\bf{k}}^2 t ^{\beta} / \Gamma (\beta + 1)\right]\label{Final}
\end{equation} 
In the real space, the fundamental solution of the fractional diffusion Eq.~(\ref{FD}) is expressible in terms of a stretched Gaussian distribution (see Ref. \cite{Rest} where further particularities of the initial conditions for time-fractional equations are discussed).

\section{Summary}
We have discussed the properties of dielectric relaxation and ac (alternating-current) conduction in disordered solids, treating them on essentially the same footing. Having assumed the property of ac universality, we found that the relaxations are stretched exponential rather than the Debye exponential. Our results comply with the classical phenomenological expressions due to Kohlrausch, Williams, and Watts (KWW). We have shown that the KWW decay function can be obtained analytically from a self-consistent model of dielectric relaxation, in which both the polarization and electric source fields are self-consistently generated by the residual charge-density. The exponent of the KWW decay function is related to the exponent of the ac conduction coefficient via $\beta = 1 - \eta$. Assuming that the conduction concentrates on a percolating cluster we found $\eta$ values within the range of observational evidence. Finally, we found that the relaxations are described by a fractional extension of the relaxation and diffusion equations, which naturally incorporate the power-law dependence of ac conduction coefficient on frequency. 

\vspace{1.0 cm}
{\bf Acknowledgments}
\vspace{1.0 cm}

A\,V\,M and K\,R gratefully acknowledge the hospitality at the University of Calabria (Italy), where the final version of this paper was written. This work was supported under the project No 171076/V30 of the Norwegian Research Council.



\end{document}